\documentclass[12pt]{article}
\usepackage{cite}
\usepackage{dcolumn}
\usepackage{bm}
\usepackage[latin1]{inputenc}
\usepackage[spanish,english]{babel}
\usepackage{amsfonts}
\usepackage{amssymb}
\usepackage{graphicx}
 \usepackage{setspace}
 \usepackage{amsmath}
 
 \usepackage[export]{adjustbox}

 \usepackage{verbatim} 
 \usepackage[normalem]{ulem}
 
\usepackage{color}
\usepackage{soul}

\usepackage[margin=3cm]{geometry}

\newcommand{\be}{\begin{equation}}
\newcommand{\ee}{\end{equation}}
\newcommand{\bea}{\begin{eqnarray}}
\newcommand{\eea}{\end{eqnarray}}

\begin{document}

\begin{titlepage}

\begin{center}
{\large \bf Event horizons are tunable factories of quantum entanglement } 
\end{center}

\begin{center}

Ivan Agullo\footnote{agullo@lsu.edu; corresponding author}\\
{\footnotesize \noindent {\it {Department of Physics and Astronomy, Louisiana State University, Baton Rouge, LA 70803, USA}}   }\\
%\affiliation{Department of Physics and Astronomy,Louisiana State University, Baton Rouge, LA 70803-4001, USA}
%
Anthony J. Brady\footnote{ajbrady4123@arizona.edu}\\
{\footnotesize \noindent {\it {Department of Electrical and Computer Engineering, University of Arizona, Tucson, 85721, USA}}   }\\
Dimitrios Kranas\footnote{dkrana1@lsu.edu}\\
{\footnotesize \noindent {\it {Department of Physics and Astronomy, Louisiana State University, Baton Rouge, LA 70803, USA}}   }\\

\end{center}

\begin{abstract}

%It is well known that event horizons generate entanglement through the  Hawking process.  We argue that the creation of entanglement can be modulated by appropriately illuminating the horizon. 
That event horizons generate quantum correlations via the Hawking effect is well known. We argue, however, that the creation of entanglement can be modulated as desired, by appropriately illuminating the horizon. We adapt techniques from quantum information theory to quantify the entanglement produced during the Hawking process and show that, while ambient thermal noise (e.g., CMB radiation) degrades it, the use of squeezed inputs can boost the non-separability between the interior and exterior regions in a controlled manner. We further apply our ideas to analog event horizons concocted in the laboratory and insist that the ability to tune the generation of entanglement
offers a promising route towards detecting quantum signatures of the elusive Hawking effect.

\end{abstract}

\vspace{1cm}

\begin{center}{\it Essay written for the Gravity Research Foundation 2022 Awards for Essays on Gravitation} \end{center}
\begin{center}Submitted on March 26, 2022 \end{center}
\end{titlepage}

The allure of black holes has captivated physicists for nearly a century, partly due to the fact that their internal mechanisms are completely concealed by a dark cloak ---the black hole's event horizon. The mystique of black holes was amplified when, in a set of seminal papers in the early 1970's \cite{Hawking74,Hawking:1975vcx}, Stephen Hawking showed that, once quantum fluctuations are accounted for, a black hole is not actually black but, instead, emits  radiation as a hot body, gradually losing its mass in what has been dubbed as the Hawking evaporation process. Even more, Hawking's calculations imply that the evaporation products are quantum mechanically entangled with the bowels of the black hole. %partner radiation falling into the interior of the black hole. 

Understanding the generation of entanglement by a black hole, with relation to its surrounding (perhaps even ``noisy") environment, elicits deeper knowledge about the black-hole evaporation phenomenon. In the past, entanglement entropy between the interior and exterior of the event horizon has been extensively used for these purposes \cite{PhysRevLett.71.3743}, but this quantity only quantifies entanglement when the global state of the system is pure. This is not the case, for instance, if a black hole is immersed in a thermal bath, e.g., the cosmic microwave background (CMB). In this essay, we leverage techniques from quantum information theory to compute and interpret the entanglement generated during the Hawking process for an evaporating black hole in a nontrivial environment. 

Specifically, we apply the theory of Gaussian states for continuous variable systems with a finite number of interacting modes \cite{serafini17QCV}. %to analyze the Hawking process. 
At first glance, our task may seem barren, since the Hawking effect is formulated in the context of field theory with its infinitely many degrees of freedom. In  Hawking's original derivation, this is manifested in the fact that  the Hawking mode reaching future null infinity (as a normalized wave-packet sharply peaked on a positive frequency mode $e^{-i\,w \, u}$), when propagated backwards in time to past null infinity, consists of a superposition of modes $e^{-i\,\tilde{w} \, v}$ over  {\em  all} frequencies $\tilde{w}$ ($u$ and $v$ are the standard retarded and advanced null coordinates, respectively). In other words, the evolution mixes \textit{infinitely} many ``in''  modes with well defined frequencies $\tilde{w}$ to produce one ``out'' mode with frequency $w$.  However, as was noticed in \cite{Wald:1975kc}, by appropriately combining ``in'' modes $e^{-i\,\tilde{w} \, v}$ with positive frequency, one can find the progenitors of the Hawking modes. These progenitors are two normalized modes $F_I(w)$ and $F_{II}(w)$ at past null infinity, which define the same ``in'' vacuum and, conveniently, have the  property that their evolution produces exactly a single ``out'' Hawking mode $e^{-i\,w \, u}$ and a single partner mode falling into the black hole. The explicit form of $F_I(w)$ and $F_{II}(w)$ is not important for our purposes and can be found in  \cite{Wald:1975kc,Wald:1995yp}.\footnote{We remark that both $F_I(w)$ and $F_{II}(w)$ are made mostly of  modes $e^{-i\,\tilde{w} \, v}$ with  ultrahigh-frequencies $\tilde{w}$ at past null infinity \cite{Wald:1975kc,fabbri05}. This is the origin of the well-known trans-Planckian problem \cite{Jacobson:1991gr,corley:1996ar}.\label{foot}} However an important observation ---not often appreciated--- is that this choice of ``in'' modes factorizes the evolution into {\em uncoupled} $w$-sectors, in the sense that pairs of modes  $F_I(w)$ and $F_{II}(w)$ do not mix with other pairs labelled by different $w$. This decoupling of  $w$-modes allows one to straightforwardly apply techniques from Gaussian quantum information theory, as we now explain.
\begin{figure}[htp]
\centering
\includegraphics[width=\textwidth]{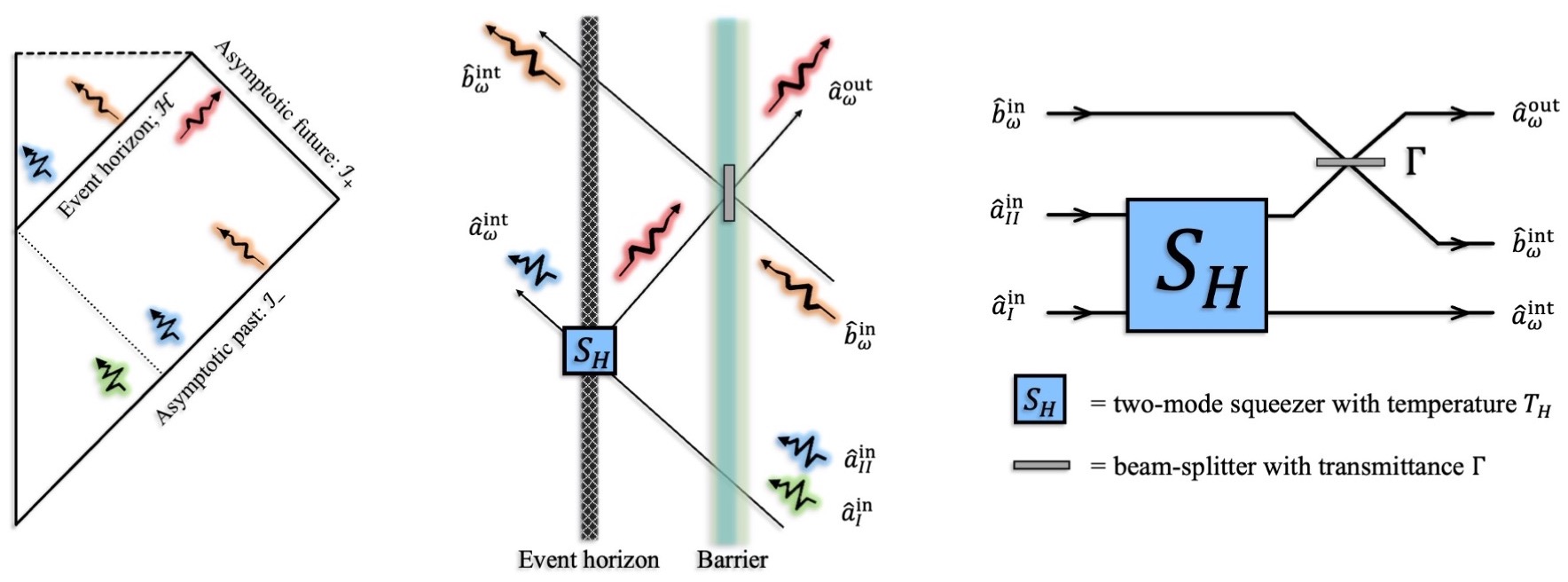}
\caption{Various depictions of the Hawking process. The main ingredients ---a two-mode squeezer associated with pair-production at the event horizon and a beam-splitter induced by the  potential barrier surrounding the horizon--- are emphasized in the middle and right panels. We illustrate the difference between low and high frequency modes by their wavelengths. The right panel represents the process in the form of a quantum circuit.}
\label{BHpic}
 \end{figure}

The evolution of the $F_I(w)$ and $F_{II}(w)$ ``in'' modes is made of two contributions of distinct physical origin, which we wish to differentiate. Let $\hat a^{\rm in}_I(w)$ and $\hat a^{\rm in}_{II}(w)$ be annihilation operators defined from normalized wave-packets sharply  peaked on the modes $F_I(w)$ and $F_{II}(w)$ at past null infinity, and $\hat a^{\rm out}_w$ and $\hat a^{\rm int}_I(w)$ be annihilation operators similarly defined from Hawking modes of frequency $w$ at future null infinity and their partner crossing the horizon, respectively. The first contribution ---which constitutes the core of the Hawking effect--- is the 
transformation
\bea \label{SQ} \hat a^{\rm in}_{I} &\to& a^{\rm int}_I\,   \cosh{r_H}- \hat a^{\rm out\, \dagger}_w \, \sinh{r_H}\, ,\nonumber  \\     \hat a^{\rm in}_{II} &\to&  \hat a^{\rm out}_w \, \cosh{r_H} -   a^{\rm int\, \dagger}_I\,  \sinh{r_H}\, , \eea 
where $r_H=\tanh^{-1}{e^{-\frac{w}{2\, T_H}}}$ and $T_H$ is Hawking's temperature. Here, we have omitted the label $w$ and the angular quantum numbers $\ell$ and $m$ for brevity. In the terminology of quantum optics, this transformation is precisely a process of {\em two-mode squeezing} \cite{gerry2005introductory}, which is responsible for the creation of entangled Hawking pairs. 

A second contribution of the Hawking process is ``back-scattering", which occurs as Hawking radiation tries to escape to infinity. In doing so, the outgoing radiation meets and scatters off the gravitational potential barrier surrounding the black hole ---leading to a portion of the wave-packet being reflected (or scattered back) into the black hole; the remaining portion gets transmitted to future null infinity. This classical scattering phenomenon involves a third ``in'' mode, made of wave-packets centered on $e^{-i \tilde{w} \, v}$ at past null infinity,  with frequency $\tilde{w} \approx w$ (i.e., back-scattering does not involve an ultrahigh blue-shift), and a second ``int''  mode falling into the horizon, centered on $e^{-i\,\tilde{w} \, v}$, also with $\tilde{w} \approx w$ (see Fig.~\ref{BHpic}). If we denote by $\hat b^{\rm in}_{w} $ and  $\hat b^{\rm int}_{w}$ the annihilation operators defined for these wave-packets, respectively, back-scattering induces the transformation,
\bea  \label{BS} \hat b^{\rm in}_{w} &\to&\hat  b^{\rm int}_{w}\,   \cos{\theta}+ \hat a^{\rm out}_w\, \sin{\theta}\, ,\nonumber  \\  
  \hat a^{\rm out}_w &\to&  \hat a^{\rm out}_w\, \cos{\theta} -  \hat b^{\rm int}_{w}\,   \sin{\theta}\, , \eea 
where $\Gamma=\cos^2\theta$ is the probability to transmit across the potential barrier, which depends on $w$, $\ell$, $m$ and the spin $s$ of the field under consideration. In the terminology of quantum optics, this is the action of a {\em beam splitter}. %It is a passive transformation, in the sense that it does not amplify modes and, therefore, leaves the vacuum invariant (no particle creation). 

Thus, for each individual frequency $w$, the Hawking process can be understood as  an evolution from three modes to three modes, $\hat a^{\rm in}_{I} ,\hat a^{\rm in}_{II} ,\hat b^{\rm in}_{w}\to  \hat a^{\rm out}_{w} ,\hat a^{\rm int}_{I}, \hat b^{\rm int}_{w} $ ---only one of which escapes to infinity--- made by concatenating a two-mode squeezer and a beam splitter, as  depicted in Fig.\ \ref{BHpic}. The beam-splitter  divides both the intensity and the entanglement generated by the squeezer, in such a way that the $\hat a^{\rm out}_{w}$ mode reaching infinity is generically entangled with {\em both}, the high frequency mode $\hat a^{\rm int}_{I},$ and the low frequency one  $\hat b^{\rm int}_{w}$   (frequencies measured by freely falling observers crossing the horizon). Describing the Hawking process in this way is advantageous, as such allows us to compute the resulting evolution of various input Gaussian states in a few simple lines. 

Recall that the information in a quantum Gaussian state is exhaustively encoded in its first moments and its covariance matrix (defined below). For each ``in'' mode, define a pair of  canonically conjugate operators $\hat x_{\alpha}=\frac{1}{\sqrt{2}}(\hat a_{\alpha}+\hat a_{\alpha}^{\dagger})$, $\hat p_{\alpha}=\frac{-i}{\sqrt{2}}(\hat a_{\alpha}-\hat a_{\alpha}^{\dagger})$, where the index ${\alpha}$ labels the three ``in'' modes. Let $\hat r^i=(\hat x_1,\hat p_1,\hat x_2,\hat p_2,\hat x_3,\hat p_3)$ be the vector of canonical operators. The first moments of a Gaussian state $\hat \rho$ are $\mu^i={\rm Tr}[\hat \rho \, \hat r^i]$, and its covariance matrix $\sigma^{ij}={\rm Tr}[ \hat \rho\, \{ \hat r^i-\mu^i, \hat r^j-\mu^j\}]$, where the curly brackets indicate an anti-commutator. 

The Hawking process is a linear evolution, as is evident from (\ref{SQ}) and (\ref{BS}); hence it preserves Gaussianity: i.e., the evolution maps an initial Gaussian state with $(\vec \mu^{\rm in}, \sigma^{\rm in})$ to another Gaussian state with $(\vec \mu^{\rm out}, \sigma^{\rm out})$. Specifying the input moments and the evolution, we can obtain any desired information about the Hawking effect. For instance, the mean number of quanta on any ``out'' mode is $\langle \hat{n}_{\alpha} \rangle=\frac{1}{4}{\rm Tr}[\sigma^{\rm out}_{\rm red}]+\vec{\mu}^{\rm out\, \top}_{\rm red}\cdot\vec{\mu}^{\rm out}_{\rm red}-\frac{1}{2}$, where  the subscript ``red'' stands for ``reduced'', and indicates the components corresponding to the concrete mode under consideration. 

Another quantity that we are particularly interested in is the entanglement between the interior and exterior regions of the black hole. Entanglement can be conveniently quantified by means of the logarithmic negativity (LogNeg) \cite{peres96, plenio05}. %LogNeg is in one-to-one correspondence with the violation of the Positivity of Partial Transpose (PPT) criterion for quantum states \cite{plenio05} ---a criterion that separable quantum states faithfully obey. 
LogNeg has important advantages for our purposes, as compared to other entanglement quantifiers or witnesses. On the one hand, it can be easily computed from  the covariance matrix $\sigma^{AB}_{\rm red}$ of any bi-partite system. On the other hand, if either subsystem A or B is made of a single mode, LogNeg is a faithful quantifier of entanglement, even for mixed states (see, e.g., \cite{serafini17QCV} for further details). We note that entanglement is encoded \textit{entirely} in the covariance matrix, with no reference to the first moments.

We apply these tools to a few examples. The simplest case is when the ``in" state consists of only vacuum fluctuations, in which case the first moments are $\vec \mu^{\rm in}_{\rm vac}=\vec 0$ and the covariance matrix is $\sigma^{\rm in}_{\rm vac}=\mathbb{I}_{6}$, where $\mathbb{I}_6$ is the $6\times 6$ identity matrix. Straightforward application of (\ref{SQ}) and (\ref{BS}) produces $\langle \hat{n}_{\rm out}(w) \rangle=\Gamma(w)\, \sinh^2 r_H(w)$ for the out mode reaching infinity, where we can distinguish the contribution from  both the  two-mode squeezer and the beam splitter. This is a well known result for the spontaneous Hawking effect \cite{Hawking:1975vcx}.

As an example of the entanglement in the final state for vacuum input, the LogNeg between the Hawking mode $a^{\rm out}_w$ and the two modes falling into the black hole is given by the following expression, which, although not particularly illuminating, shows that this quantity can be analytically expressed in closed form
\begin{align*}  &{\rm LogNeg}[\hat a^{\rm out}_w|(\hat a^{\rm int}_I,\hat b^{\rm int}_w)]  = {\rm Max}\Bigg[0,\\& -{\rm Log}_2\Big[\frac{1}{\sqrt{2}} \Big(\Big| -2 \cos 4 \theta \, \sinh ^4r_H+\cosh ^2r_H  -2 \cos 2 \theta \,  \sinh ^22 r_H  -3 \cosh ^2r_H\,  \cosh 2 r_H\\ & +\sqrt{-4 + \big(2 \cos 4 \theta \,  \sinh ^4r_H+\cosh ^2r_H \, (-1 +3 \cosh 2 r_H+ 8  \cos 2 \theta \,  \sinh ^2r_H)\big)^2}\Big| \Big)^{1/2}\Big]\Bigg]. \end{align*}
By plotting this expression, one can check that the presence of the potential barrier degrades the entanglement carried out to infinity. For example, entanglement vanishes in the limit $\theta \to \pi/2$ ($\Gamma\to 0$), since, in that limit, the barrier completely blocks the outgoing Hawking radiation. This in turn implies that, for a fixed frequency $w$, modes with the lowest angular multipoles $\ell$ carry most of the entanglement (as well as most of the energy).

If we replace the initial vacuum by an excited state, we are in the realm of the stimulated Hawking process. A simple scenario here is when the initial state is a coherent state. Coherent states are ``displaced vacua'', in the sense that they are Gaussian states with the same covariance matrix as the vacuum, but have different first moments: $(\vec \mu_{\rm coh}\neq \vec 0,\,  \sigma_{\rm coh}=\mathbb{I}_6)$. Using the expression  for $\langle \hat{n}_{\alpha} \rangle$ given above, we quickly see that the mean number of quanta in the Hawking mode $a^{\rm out}_w$ gets amplified, as expected. However, the entanglement structure in the final state remains {\em exactly the same} as for vacuum input, as the covariance matrices for each are identical. Hence, seeding the process with coherent states does not change quantum aspects of the final state. This justifies the common lore about the intrinsically classical character of the stimulated Hawking effect. We now argue, contrary to the common lore, that such is not true for other input states. 

\begin{figure}[t]
\centering
\includegraphics[width=.485\linewidth,,valign=c]{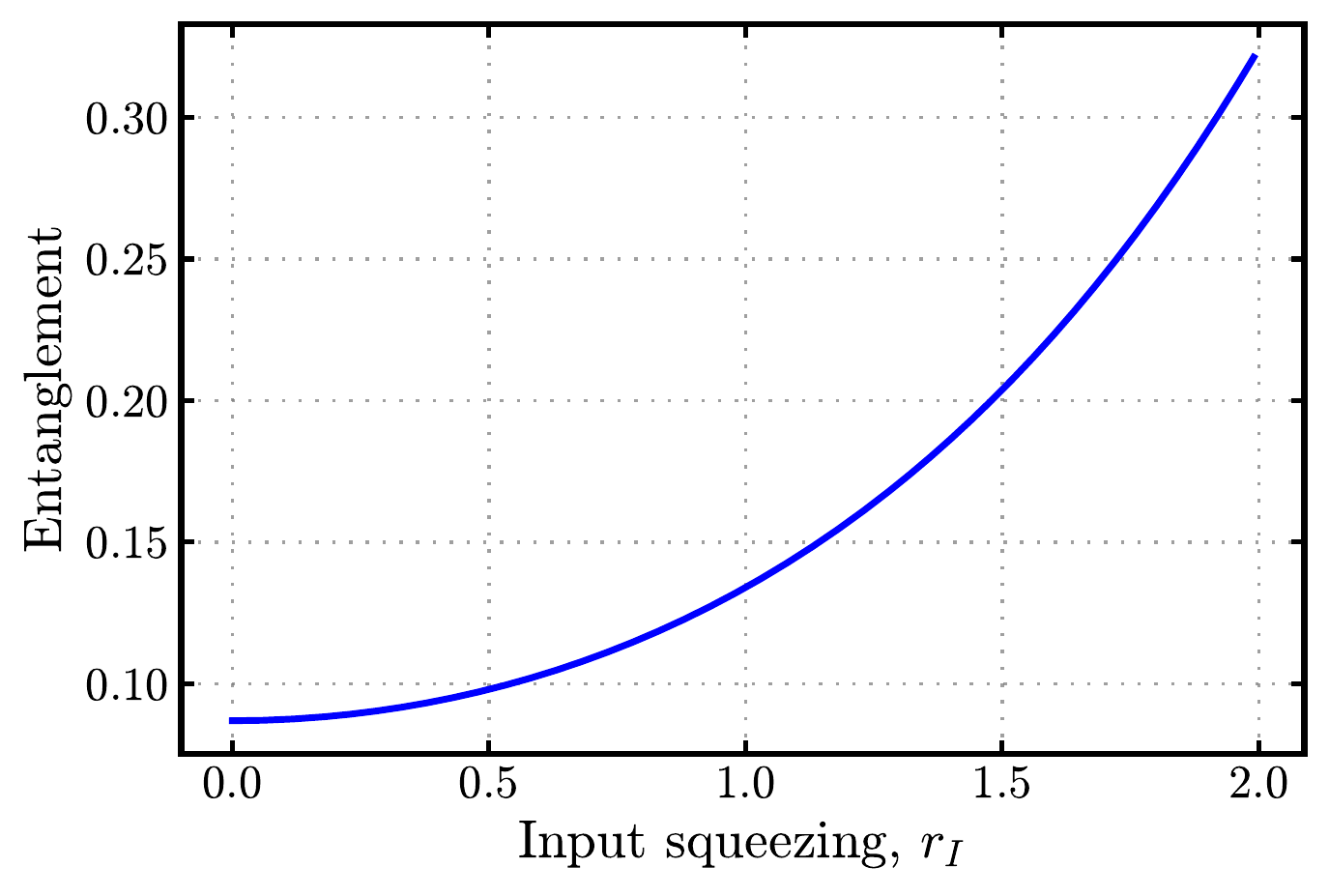}\quad \includegraphics[width=.485\linewidth,,valign=c]{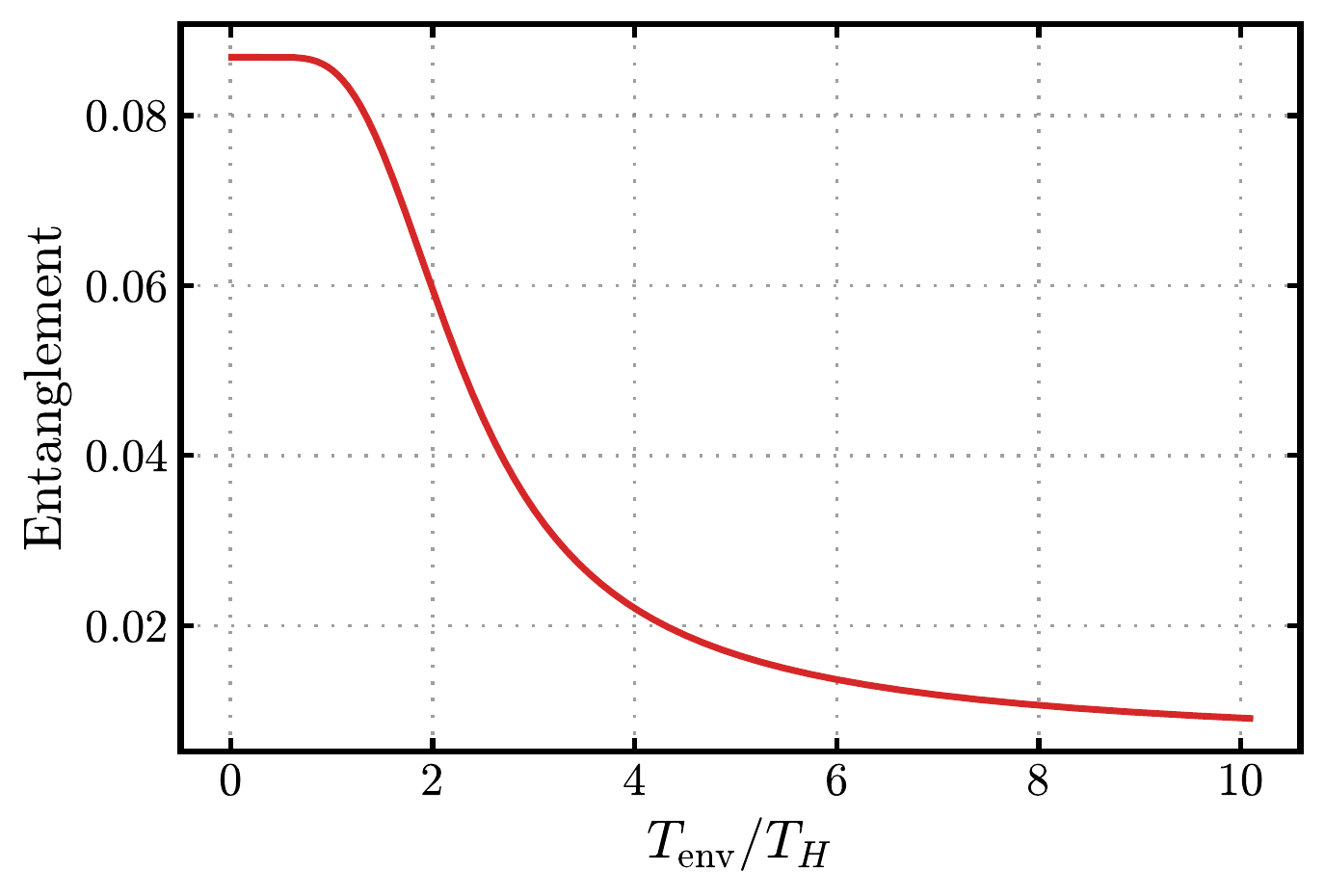}
\caption{Left Panel:  ${\rm LogNeg}[\hat a^{\rm out}_w|(\hat a^{\rm int}_I,\hat b^{\rm int}_w)]$ measuring the entanglement between the Hawking mode  reaching infinity and the  interior of a Schwarzschild black hole, for photons with $w=6.25\, T_H$, $\ell=1$, $m=0$ (the mode which emits the maximum amount of energy in electromagnetic radiation, and for which  $\Gamma=0.4689$ \cite{Page:1976df}) as a function of the squeezing intensity $r_I$ in the ``in'' mode $\hat a^{\rm in}_I$. Right Panel:  Same quantity versus the environment temperature (no initial squeezing; $r_I=0$).}
\label{LogNeg}
 \end{figure}

By illuminating the black hole with, e.g., a single-mode squeezed state in either of the modes $\hat a^{\rm in}_I$ or  $\hat a^{\rm in}_{II}$, we find that both the number of final quanta in the Hawking modes $\hat a^{\rm out}_{w}$ and its entanglement with the interior modes are amplified. Consider a squeezed vacuum in the $\hat a^{\rm in}_I$ mode,  $(\vec \mu_{\rm sqz}= \vec 0,\,  \sigma_{\rm sqz}= e^{\sigma_z\, 2 \, r_I} \,\oplus \mathbb{I}_4)$, where $\sigma_z$ is the familiar $z$-Pauli matrix and $r_I$ is the initial squeezing intensity. Fig.~\ref{LogNeg} shows that the entanglement between the black hole interior and the exterior (formally quantified by ${\rm LogNeg}[a^{\rm out}_w|( a^{\rm int}_I, b^{\rm int}_w)]$), for a given frequency $w$, grows with  $r_I$. Note that a single-mode squeezed state does not contain any entanglement among each of the three ``in'' modes, so the final entanglement is entirely generated during pair-production in the Hawking process. Hence, by tuning the initial squeezing $r_I$, one can enhance the entanglement generated by black holes. There are multiple reasons that make the implementation of this idea impossible for astrophysical black holes; the most prominent being that the two ``in'' modes $\hat a^{\rm in}_I$ and $\hat a^{\rm in}_{II}$ are made of ultrahigh-frequency modes at past null infinity. However, as we argue below, the implementation in analog event horizons produced in the lab is feasible with present technology, since the blue-shift between frequencies of  ``in'' and ``out'' modes is not exponentially large, as in the astrophysical case \cite{philbin08}. 

Let us next consider thermal  input radiation. This is of obvious interest since all astrophysical black holes are immersed in the CMB. Thermal states are mixed Gaussian states, with zero mean, no correlations among modes, and covariance matrix for each individual mode equal to $(2\, n^{\alpha}_{env} +1)\, \mathbb{I}_2$, where $n^{\alpha}_{\rm env}$ is the number of thermal quanta in the mode $\alpha$. Since the modes $\hat a^{\rm in}_I$ or $\hat a^{\rm in}_{II}$ are ultrahigh-frequency modes (see footnote \ref{foot}), and assuming the temperature is small compared to the Planck scale, it is reasonable to substitute  $n_{env}\approx0$ for them. The relevant thermal quanta in the initial state occupy only the low frequency mode $\hat b^{\rm in}_w$. The initial state is thus characterized by $\vec \mu^{\rm in}_{\rm th}=\vec 0$ and  $\sigma_{\rm th}\approx \mathbb{I}_4\oplus (2\, n_{\rm env}+1)\, \mathbb{I}_2$, from which we find $\langle \hat{n}_{\rm out} \rangle= n_{\rm env}\, (1-\Gamma)+\, \Gamma\, \sinh^2{r_H}$ quanta reaching infinity. This is the same result that we would obtain for vacuum by simply adding  the thermal quanta scattered back to infinity by the potential barrier. From here, we see that if $n_{\rm env}$ is equal to $\sinh^2r_H$ (i.e.,  $T_H=T_{env}$) we have $\langle \hat{n}_{\rm out} \rangle- n_{\rm env}=0$ for all $w$, and the black hole is in equilibrium, as expected from thermodynamical arguments. Note that the black hole loses mass only if $T_H>T_{\rm env}.$

It is intriguing to wonder how the entanglement between the interior and exterior regions of the black hole is modified when the black hole is immersed in a thermal bath; so we compute the entanglement in this scenario and plot the results in Fig.~\ref{LogNeg}. We see that thermal fluctuations of the bath are detrimental to the quantum entanglement between the emitted Hawking quanta and the black hole. Hence, although the mode $\hat b^{\rm in}_w$ is only involved in the process of back-scattering, thermal fluctuations present in this mode degrade the entanglement produced in the Hawking process. A physical consequence of this is that, since astrophysical black holes have Hawking temperatures several orders of magnitude lower than the CMB temperature, the faint Hawking quanta they emit are barely entangled with the black hole's interior. This can be seen quantitatively in the right panel of Fig.~\ref{LogNeg}, where we observe a sharp transition in the entanglement as $T_{\rm env}\gtrsim T_{H}$.  Remarkably though, for arbitrarily high environmental temperatures, the entanglement saturates to a small ---but still \textit{non-zero}--- value. (We note that astrophysical black holes are also immersed in noisy backgrounds of other fields that contribute significantly to the Hawking process, like the cosmic background of neutrinos and the stochastic background of gravitational waves. Both will generically degrade the generation of entanglement for these fields, especially for neutrinos due to their fermionic character.)

We finish this essay by arguing that the ideas thus presented have direct applicability for analog event horizons manufactured in the laboratory. We highlight this with a popular scenario to recreate the physics of the Hawking process: optical systems \cite{philbin08,demircan11TRANSISTOR,rubino2012soliton,petev2013blackbody,finazzi13,Belgiorno:2014ana,linder16,Bermudez:2016hbl,Belgiorno:2017glw,drori19,jacquet20emission,Jacquet:2020jpj,rosenberg2020optical,Aguero-Santacruz:2020krw}. In an optical analog, an electromagnetic pulse in a dielectric  material locally modifies the refractive index via the so-called Kerr effect. Hence, by introducing a strong pulse in a dielectric material, the speed of weak probes propagating thereon can be tuned. Probes that are initially faster than the  pulse will slow down when trying to overtake it, and if the pulse is strong enough, its rear end acts as an (moving) impenetrable barrier. This is the  analog of the horizon of a white hole. Similarly, an analog black hole horizon appears in the front end of the pulse. The creation of these white-black holes has been experimentally carried out, and the stimulated radiation with coherent inputs has been recently demonstrated in \cite{drori19}. Optical systems offer a great advantage to generate, manipulate, and observe quantum states as well as their entanglement structure \cite{raymer2009} ---tasks that are routinely done with present technology.

\begin{figure}[htp]
\centering
\includegraphics[width=.5\textwidth,valign=c]{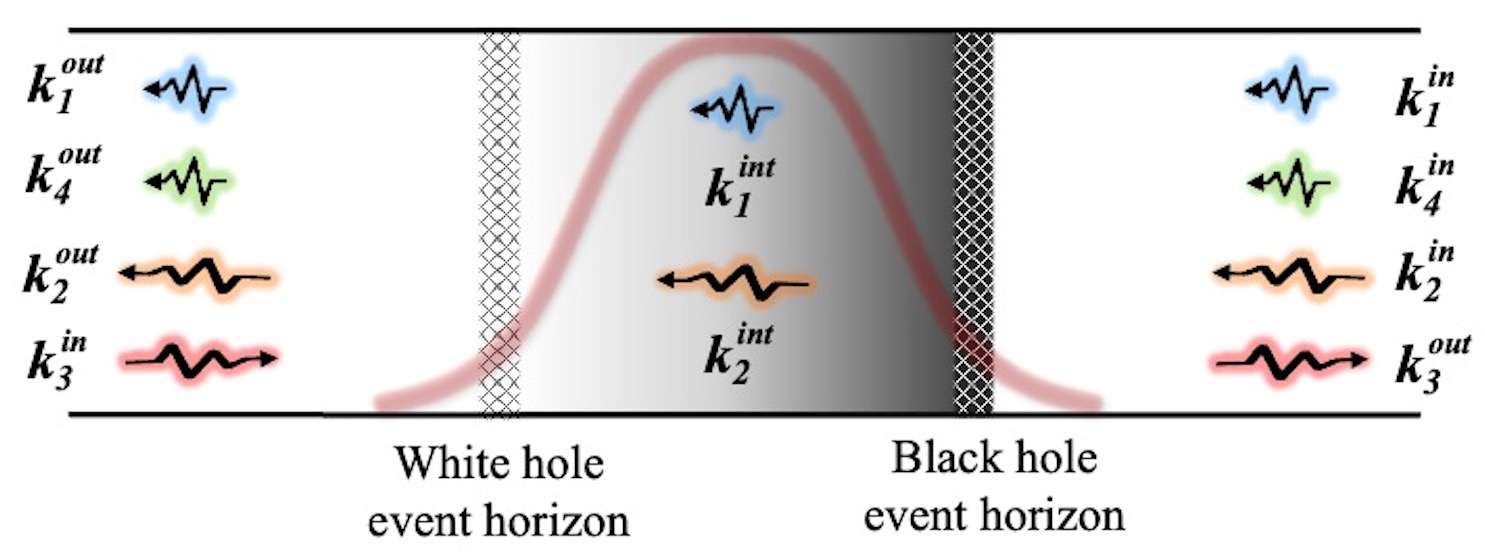}
\includegraphics[width=.47\textwidth,valign=c]{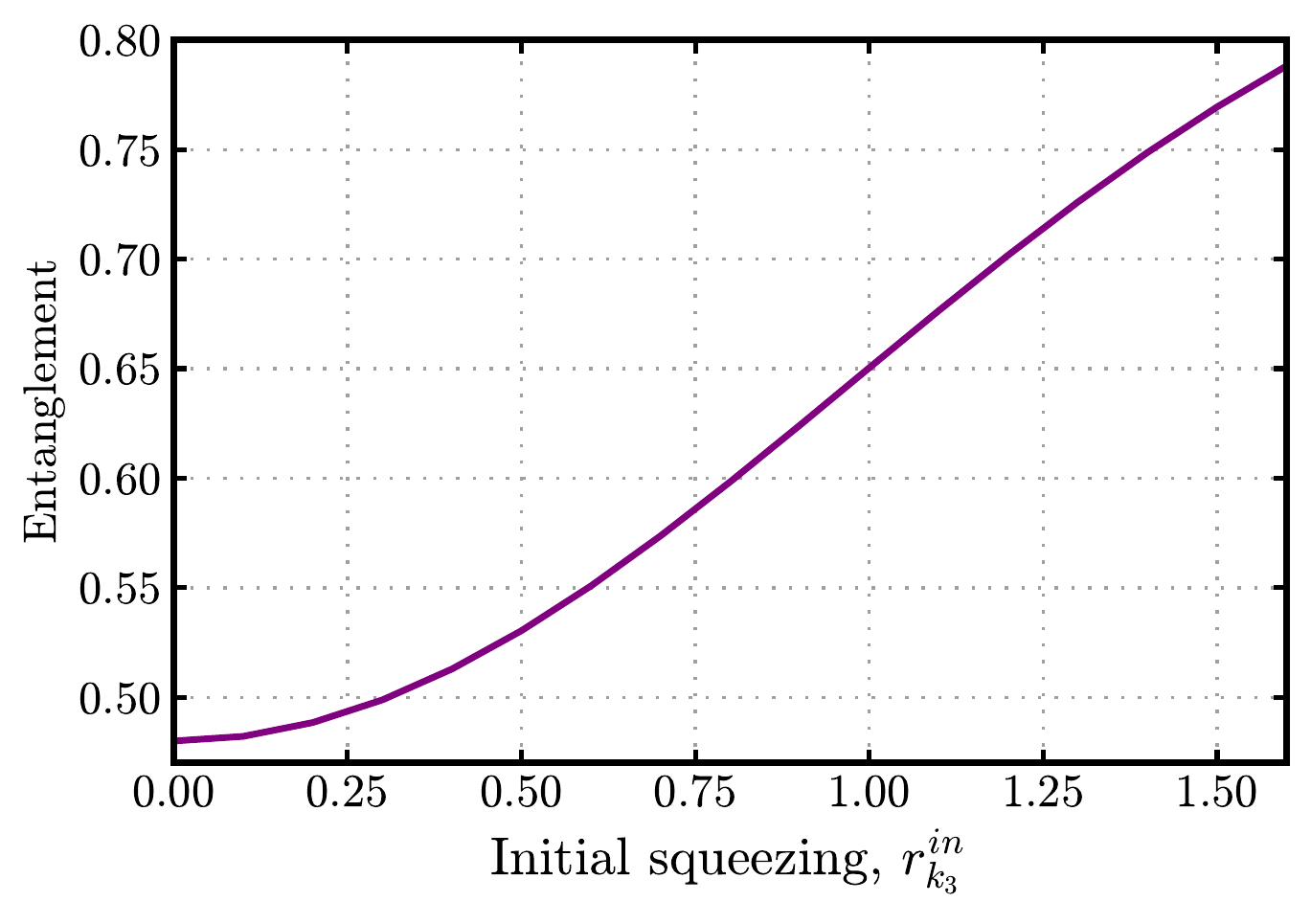}
\caption{Left panel: Illustration of the structure of modes for an optical analog white-black hole in the frame comoving with the pulse; the modes $k_i$ shown are the solutions to the dispersion relation $k(w)$ for a fixed frequency $w$. %The lab frame wave-lenghts are (in nm) $\lambda_1=233.7$, $\lambda_2=3.47\times 10^{6}$, $\lambda_3=28784.2$, $\lambda_4=235.633$. 
Right Panel: LogNeg between the outgoing white-hole Hawking-pair  $k^{\rm out}_1$ and $k^{\rm out}_4$ versus the  initial squeezing intensity $r_{k_3}^{\rm in}$, for $w=2T_H$. We assume isotropic thermal noise, $\bar n_{\rm env}=(e^{w/T_{\rm env}}-1)^{-1}$ with ${T_{\rm env}=T_H}$, and $10\%$ losses. %Measuring LogNeg for different values of the initial squeezing $r_{k_3}^{\rm in}$ permits to identify whether the observed entanglement is originated in the Hawking process and to measure  $r_H(w)$, which in turns allows to test the black body spectrum.
}\label{modestructure}
 \end{figure}

 The decomposition of the Hawking process in terms of two-mode squeezers and beam-splitters, as described in this essay, can be repeated for these optical systems \cite{Agullo:2021vwj}, and although the scenario is richer than that of astrophysical black holes, due to non-trivial dispersion relations of the media, the conclusions are similar. Namely, the generation of entanglement in the Hawking process can be tuned, either degrading it by illuminating the system with noisy thermal photons or amplifying it using squeezed inputs. Some quantitative results of our recent detailed analysis  \cite{Agullo:2021vwj} are shown in the right panel of Fig.~\ref{modestructure}. These calculations  incorporate the effects of ambient noise and losses, both ubiquitous in real experiments. The ability of tuning the input quantum state provides a sharp tool to test the two defining aspects of the Hawking process, namely the generation of quanta with a black body distribution and a concrete entanglement structure, and a  protocol to observe these features has been put forth in \cite{Agullo:2021vwj}. \\

\noindent{\bf{\em Acknowledgments.}}
We have benefited from  discussions with A. Ashtekar, D. Bermudez, M. Jacquet,  A. Fabbri, J. Pullin, J. Olmedo, O. Magana-Loaiza and J. Wilson. We thank A. del Rio for assistance with the evaluation of the grey body factors.  This work is supported by the NSF grant PHY-2110273, and by the Hearne Institute for Theoretical Physics. A.J.B. also acknowledges support from the Defense Advanced Research Projects Agency (DARPA) under Young Faculty Award (YFA) Grant No. N660012014029.

\bibliographystyle{utphys}

\providecommand{\href}[2]{#2}\begingroup\raggedright\endgroup

\end{document}